\documentclass[pra,twocolumn]{revtex4-2}

\usepackage{epsfig,amsmath}
\usepackage{subfigure}
\usepackage{graphicx}
\usepackage{dcolumn}
\usepackage{stmaryrd}
\usepackage{mathrsfs}
\usepackage{pifont}
\usepackage{amsthm}
\usepackage{amssymb}
\usepackage{bm}
\usepackage{latexsym}
\usepackage{color}
\usepackage{hyperref}
\usepackage{multirow}
\usepackage{xcolor}
\usepackage{babel}
\usepackage{verbatim}
\usepackage{enumerate}
\usepackage{placeins} 
\usepackage{booktabs}
\usepackage{dsfont}
\usepackage{amssymb}
\usepackage{algorithm}
\usepackage{algpseudocode}

\begin{document}

\title{Optimal control of open-quantum-system dynamics predicted by long short-term memory}
                                                                                                   
\author{Jun-Dong Zhong\textsuperscript{1}, Zhao-Ming Wang\textsuperscript{1,2,3}}
\email{Contact author: wangzhaoming@ouc.edu.cn}

\address{$^{1}$College of Physics and Optoelectronic Engineering, Ocean University of China, Qingdao 266100, China}
\address{$^{2}$Engineering Research Center of Advanced Marine Physical Instruments and Equipment of Ministry of Education, Qingdao 266100, China}
\address{$^{3}$Qingdao Key Laboratory of Advanced  Optoelectronics, Qingdao 266100, China}
\date{\today}
\begin{abstract}
The realization of high-fidelity quantum control is crucial for quantum information processing, particularly in noisy environments where control strategies must simultaneously achieve precise manipulation and effective noise suppression. Conventional optimal control designs typically require numerical calculations of the system dynamics. Recent studies have demonstrated that long short-term memory neural networks (LSTM) can accurately predict the time evolution of open quantum systems. Based on LSTM predicted dynamics, we propose an optimal control framework for rapid and efficient optimal control design in open quantum systems. As illustrative examples, we apply the proposed framework to design optimal control for adiabatic speedup in a two-level system and for quantum state transfer in a spin chain, both under non-Markovian environments. For adiabatic speedup, our optimization procedure involves two steps: driving trajectory optimization and zero-area pulse optimization. Fidelity improvements for both steps have been obtained, demonstrating the effectiveness of the scheme. Furthermore, this effectiveness is validated for quantum state transfer in a spin chain, which is a high dimensional control problem. Our optimal control design scheme utilizes predicted dynamics to generate optimized controls, offering broad application potential in quantum computing, communication, and sensing. 
\end{abstract}

\maketitle
\section{Introduction}

High-precision and robust quantum control is an essential prerequisite in performing high-quality quantum information processing tasks. 
Dynamical decoupling \cite{Kovacs1999, Yang2024}, as one of the effective control strategies, has been widely utilized in different experimental platforms, such as superconducting quantum circuit \cite{Tripathi2025,Evert2025}, Rydberg atoms \cite{Ocola2024,Bluvstein2021,Li2026} and trapped ions \cite{Biercuk2009,Morong2023,Piltz2016}. The realization of the high-fidelity control depends on the design of the pulse profile, e.g., optimal radio frequency pulse sequence in superconducting quantum circuits \cite{Zhang2025,9286506,Guoguoping24}. Both the analytical and numerical methods have been developed to design the pulse sequence \cite{wang2020almost,VanDamme2017}. For the analytical methods, zero-area pulse scheme \cite{wang2020almost,Liyiye2025,Turinici2019} have been proposed, which can be used to prevent the transition of the states from one subspace to other subspaces. The global optimal pulses for the control of two-level quantum systems against offset or control-field uncertainties are derived via the Pontryagin maximum principle \cite{VanDamme2017}. The fastest possible pulses that implement single-qubit phase gates via geometrical optimization have been investigated \cite{Zengjunkai2018}. In addition, numerous traditional numerical methods have been proposed for pulse design, including stochastic gradient
descent and Adam algorithm \cite{Turinici2019}, Gradient Ascent Pulse Engineering \cite{Khaneja2005}, the Krotov method \cite{Goerz2019}, the Chopped Random Basis method \cite{Mller2022} and the distributed proximal policy optimization algorithm \cite{Lu2025}, etc. Furthermore, recently there has been a surge in artificial intelligence-driven optimal control design, such as techniques encompassing reinforcement learning \cite{LiShengyong2025} and supervised learning \cite{AbuNada2025}, among others.

In practice, environmental noise will degrade the control performance \cite{https://doi.org/10.48550/arxiv.2408.09637}. And it is generally difficult to design optimal controls in a noisy environment, especially for the analytical methods \cite{Chenzijie2025}.  Lindblad quantum master equation has been widely used to describe the dynamics in the presence of environment \cite{Campaioli2024}. The non-Markovian cases for bosonic baths has been derived via quantum state diffusion equation approach \cite{Chenyusui2015} for both zero temperature \cite{Strunz1999} and finite temperature \cite{Wangjpa2021}. The influence of the baths on the system can be characterized by environmental parameters such as system-bath coupling strength, bath temperature, and bath characteristic frequency. Therefore, based on the master equation, one common pulse-design approach is to use the ideal pulses derived in the absence of environment as a foundation, followed by a fine-tuning to mitigate the adverse effects of noise \cite{Xieyangyang2025,Xie2022,Chenzijie2025}. However, during the design process, it is necessary to continuously solve the master equation through numerical methods, including finite difference, Runge-Kutta, Low-Storage Runge-Kutta \cite{Yan2017}, etc. The drawbacks of these methods are two-fold: one is slow speed, and the other is the difficulty in dealing with the large Hilbert space dimension.  

\begin{figure*}[htbp]
	\centering
	\includegraphics[width=\linewidth]{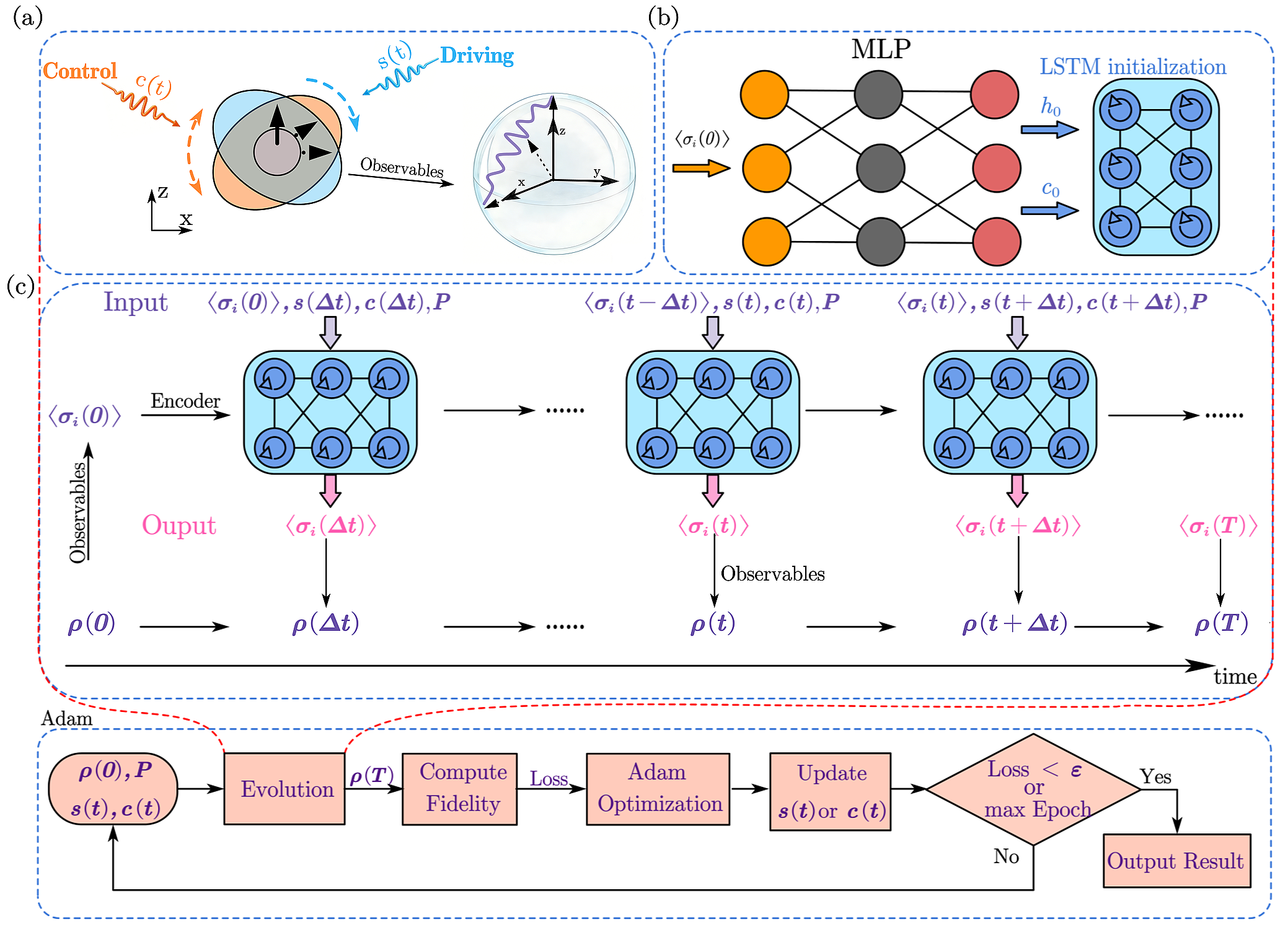}
	\caption{Schematic illustration of the machine learning algorithm for open quantum system dynamics prediction and control optimization. Top left: The predicted evolution of a two-level system (e.g., a spin) under driving $s(t)$ and control $c(t)$. Top right: The expectation values of the initial states are encoded into the initial hidden states of the LSTM. The initial state is first encoded into the LSTM hidden state via a multilayer perceptron (MLP). This encoded state is then used by the network to autoregressively generate the system’s evolution trajectory across the entire time interval, given the control pulses and environmental parameters. Middle: Detailed architecture of the LSTM, highlighting the input–output relations between the quantum system and the network. Bottom: The process of pulse optimization via the Adam algorithm. The dynamics of the system are predicted through the LSTM. The fidelity of the final state is taken as the loss function.}
	\label{Fig.1}
\end{figure*}

Recently, machine learning based NNs has been widely applied in simulating the open quantum system dynamics, which can partially
address the aforementioned challenges. Various architectures have been explored, such as physics-informed neural networks \cite{LiZhenyu2025}, deep quantum neural network \cite{Pan2023}, variation methods \cite{Guo2025, Bond2024}, autoregressive NN \cite{LuoDi2022}, time series prediction \cite{https://doi.org/10.48550/arxiv.2401.06380}, and recurrent neural networks (RNNs) \cite{Banchi2018}. The intuition behind these approaches is that, if there exists an efficient description of the system, this can be learnt from data without using explicit modelling assumptions. In particular, NN-based learning techniques have shown that it is possible to learn complex functional dependencies in time series directly from the data, without trying to make theoretical assumptions that may be unjustified (such as the weak coupling and high temperature approximation in deriving non-Markovian quantum master equation \cite{Wangjpa2021}) or hard to derive from phenomenological observations. LSTM, a type of RNNs specifically designed to model dynamical systems with possibly long range temporal correlations, is well-suited for predicting non-Markovian dynamics. It can quickly learn the underlying correlations in time to correctly predict the evolution of open quantum system displayed in the data \cite{An2025}. The runtime of the LSTM is shorter than direct numerical simulations \cite{Mohseni2022}. Furthermore, the LSTM are able to predict the dynamics very well for the time interval that it has been trained on, and also to extrapolate to a longer time window that it has not been trained.

As the LSTM has memory, it is able to build up by itself an implicit representation of the higher-order correlation using a non-Markovian strategy \cite{Mohseni2022}. Owing to their efficiency, in the optimal pulse design, we employ LSTM to predict system dynamics, eliminating the need for direct numerical calculation. The LSTM plays the role of a solver of the master equation. This prediction is then integrated with conventional optimization algorithms for the optimal pulse design. The advantage is that it can output the pulse design quickly compared with the numerical calculation. 
In the quantum setting, RNNs have been previously employed for quantum control \cite{August2017,Ostaszewski2019}. In this paper, we consider the optimal control design in the adiabatic speedup as an example. Adiabatic speedup allows quick evolution into target states of otherwise slow adiabatic dynamics, and several schemes have been suggested, including  transitionless quantum driving \cite{Lohe2008}, counteradiabatic control \cite{epait2023}, the rapid adiabatic passage \cite{LiKehui2023} and stimulated Raman adiabatic passage \cite{Kumar2016}, etc. 
Recently, schemes for accelerating the adiabatic process via zero-area pulse were investigated \cite{Xieyangyang2025}. The optimal pulse obtained via stochastic learning \cite{Xie2022} allows to acquire higher adiabatic fidelities than in the ideal pulses. In this paper, we use the LSTM and the Adam algorithm together for the optimal design of both driving trajectory and zero-area pulse. The system dynamics are predicted by the LSTM instead of numerical calculation using Runge-Kutta \cite{Xie2022}. Our strategy substantially improves adiabatic fidelity, demonstrating its rapid and efficient performance in optimal control design. Furthermore, we validate the feasibility of the proposed scheme in high-dimensional systems by applying it to the optimal control of quantum state transfer in an $XY$ spin chain. 
The advantage of our proposed algorithm (neural-network-based dynamics prediction combined with optimization) is that it offers a scalable alternative, avoiding the rapid scaling of computational cost with Hilbert space dimension that is inherent to numerical propagation-based optimal control.

\section{Model}
	
For the adiabatic evolution, a time-dependent Hamiltonian $H(t)$ can be written as \cite{Farhi2001, Albash2018}  
\begin{equation}
H(t) = [1-s(t)]H_i+s(t)H_f,
\label{eq:1}
\end{equation}
where $s(t)$ is the driving trajectory. $H_{i}$ and $H_f$ denote the initial and final Hamiltonians, respectively. By setting $s(0)=0$ and $s(T_{tot})=1$, we focus on the problem of obtaining a ground state at $s(T_{tot})$ by evolving the system from a trivial ground state at $s(0)=0$. Here $T_{tot}$ is the total evolution time. With the computational problem encoded in the ground state of $H_f$, adiabatic quantum computation could be realized.

For a closed system, adiabatic speedup can be achieved by adding an leakage elimination operator \cite{Wang201803,Markaida2020} Hamiltonian to the system, which effectively suppresses transitions from one subspace to the other subspaces. The system Hamiltonian becomes \cite{wang2020almost}
\begin{equation}
H(t) = [1+c(t)][(1-s(t))H_i+s(t)H_f],
\end{equation}
where \(c(t)\) denotes the pulse control function. The ideal pulse control conditions have been theoretically derived via P-Q partitioning techniques \cite{Wu2022,Wangzhaoming2020}. Various pulse shapes (e.g., rectangular, sine pulses) have been considered. For rectangular case, the idea pulse condition is $I\tau =2k\pi, k=1,2,3,... $ \cite{wang2020almost}. For the sine pulse $c(t)=I sin(\pi t/\tau)$, the ideal control conditions satisfy  $J_0(I\tau/\pi)=0$ \cite{wang2020almost}. Here $I$ is the pulse intensity and $\tau$ is the pulse half period, $J_0(x)$ is the zero-order Bessel function of the first kind. 

Now suppose the system (adiabatic channel) is immersed in bosonic baths, which are ubiquitous in solid-state quantum systems, such as quantum dots \cite{brandes2002adiabatic}, superconducting qubits \cite{devoret2013superconducting}, and NV centers \cite{balasubramanian2009ultralong}. A non-Markovian quantum master equation has been used to describe the dynamics of the system \cite{Wangjpa2021}
\begin{equation}
\begin{aligned}
\frac{\partial \rho}{\partial t} = & -i[H, \rho] + [L, \rho \overline{O}_z^{\dagger}] - [L^{\dagger}, \overline{O}_z \rho] \\
& +  [L^{\dagger}, \rho \overline{O}_w^{\dagger}] - [L,  \overline{O}_w \rho] ,
\end{aligned} 
\label{eq:3}
\end{equation}
where the operator $\overline{O}_{z,(w)}(t)$ is an Ansatz and defined as $\overline{O}_{z,(w)}(t) = \int_0^t ds \alpha_{z,(w)}(t-s) O_{z,(w)}(t,s)$. For details, see Refs.~\cite{diosi1998non,Wangjpa2021}. Consider the Lorentz-Drude spectrum as an example, the spectral density is given by $J(\omega) = \frac{\Gamma}{\pi} \frac{\omega}{1 + \left(\omega/\gamma\right)^2}$
\cite{wang2010coherent,ritschel2014analytic,meier1999non}.
 Here, $\Gamma$ represents the coupling strength between the system and the environment, while $1/\gamma$ provides a scale for the environmental memory time.  $T$ denotes the bath temperature. For non-Markovian baths, $\overline{O}_z$ and $\overline{O}_w$ satisfy \cite{Wangjpa2021}
\begin{equation}
\begin{aligned}
\frac{\partial \overline{O}_z}{\partial t} = & \left( \frac{\Gamma T \gamma}{2} - i \frac{\Gamma \gamma^2}{2} \right) L- \gamma \overline{O}_z  \\
&+ \left[-iH - (L^\dagger \overline{O}_z + L \overline{O}_w) , \overline{O}_z \right],
\end{aligned}
\label{eq:4}
\end{equation}
\begin{equation}
\begin{aligned}
\frac{\partial \overline{O}_w}{\partial t} = & \frac{\Gamma T \gamma}{2} L^\dagger - \gamma \overline{O}_w \\
& + \left[ -iH - (L^\dagger \overline{O}_z + L \overline{O}_w) , \overline{O}_w \right].
\end{aligned}
\label{eq:5}
\end{equation}
Using Eq.~(\ref{eq:3})-Eq.~(\ref{eq:5}), the non-Markovian dynamics can be fully analyzed.

In this paper, by combining LSTM and Adam algorithm, we will first focus on the optimal choice of time sequence $s(t)$. Then we concentrate on the global optimal control design $c(t)$ in the adiabatic speedup (Fig.~\ref{Fig.1}(a)).  For both scenarios, we consider the cases with and without an environment, as environmental noises typically disrupt adiabatic evolution. Now for simplicity, we consider a two-level system as an example \cite{Wu2022,Xie2022}. Take $H_i=\sigma_z$, $H_f=\sigma_x$. Here $\sigma_{z,(x)}$ is the Pauli matrix of the $z,(x)$ component. The initial state of the system is taken as the ground state of $\sigma_z$, i.e., $\left \vert E_0(0) \right \rangle= \left \vert 0\right \rangle$. According to the adiabatic theorem, if $T_{tot}$ is infinitely long, the system will evolve along an adiabatic path. The final state will be $\left \vert E_0(T_{tot}) \right \rangle=\frac{1}{\sqrt 2}(\left \vert 0\right \rangle-\left \vert 1\right \rangle)$. When the system is subject to a non-Markovian, finite-temperature heat bath, the fidelity $F = \sqrt{\left\langle E_0(t)\right \vert \rho(t) \left \vert E_0(t)\right\rangle}$ can be used to measure the adiabaticity. $\rho(t)$ is the reduced density matrix of the system in Eq.~(\ref{eq:3}) and $\left\vert E_0(t)\right\rangle $ is the instantaneous ground state of the Hamiltonian in Eq.~(\ref{eq:1}). As an example, the Lindblad operator is taken as $L = \sigma_-$ throughout.

\subsection{LSTM for the prediction of open quantum system dynamics}

\begin{figure*}[htbp]
  \centering
    
      \includegraphics[width=\linewidth]{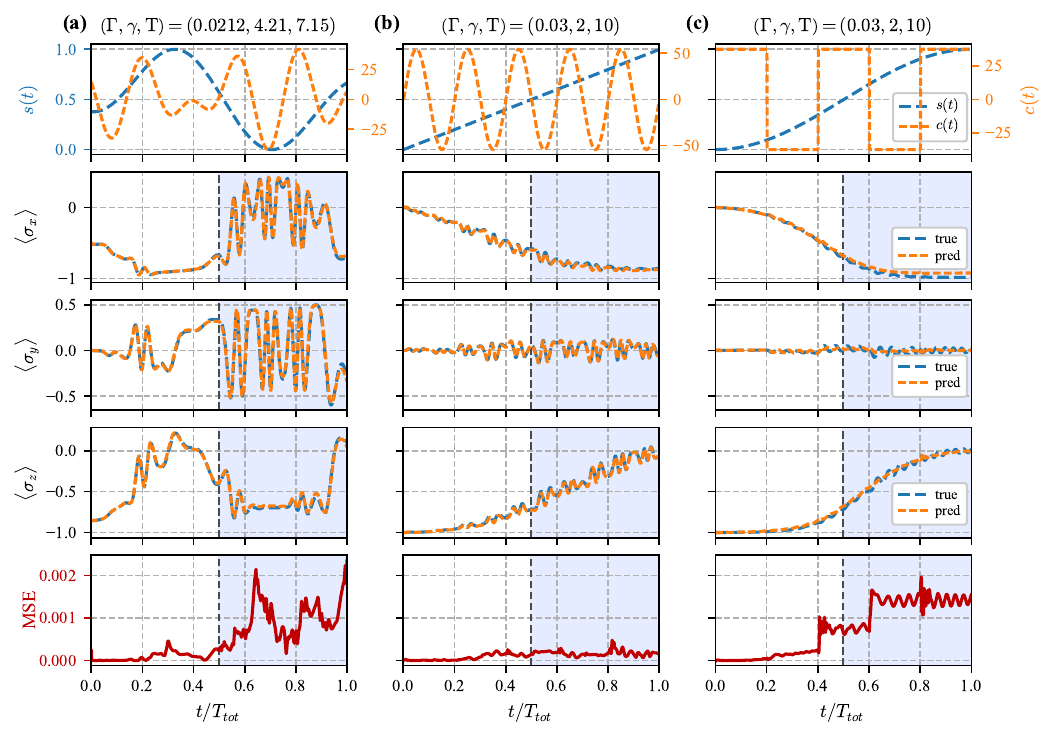}
      \caption{Verification of the high accuracy prediction ability of the LSTM for the system dynamics under arbitrary driving and control pulses $s(t)$ and $c(t)$, and arbitrary environmental parameters $\Gamma \in [0,0.05]$, $\gamma \in [1,30]$, $T\in [5,15]$. We train the NN on random driving generated via multi-frequency random sampling method based on Fourier transform for a fixed time interval $T_{tot}=2.5$ and predict the dynamics for arbitrary driving field in a large time interval $T_{tot}=5$: (a) Random Fourier Synthesis (RFS) driving and control, (b) linear ($s(t)=t/T_{tot}$) and sine ($c(t)=I\sin(2\pi t)$ ($I\approx54.4\ \text{to satisfy the third zero of } J_0(I\tau/\pi))$), (c) sine ($s(t)=\frac{1}{2}\sin(\pi t/T_{tot}-\frac{\pi}{2})+\frac{1}{2}$) and quench ($\tau=2,I\approx37.7 \ \text{to satisfy}\ \ I\tau = 2k\pi\ \text{where}\ \ k=6$ ). The initial states for three cases are the ground states of the time-dependent Hamiltonian $H(t)$ at $t=0$. These three drivings are not seen by the NN during training. The time window $[2.5, 5]$ highlighted in light purple corresponds to the region where the network was not trained. The first line is the driving $c(t)$ and $s(t)$. The 2-4th line are the dynamics of $\langle \sigma_x \rangle$,$\langle \sigma_y \rangle$, and $\langle \sigma_z \rangle$, respectively. The last line shows the evolution of the error of the NN.}
\label{Fig.2}
\end{figure*}

LSTM is a specialized type of RNNs, specifically designed to model dynamical systems with long-term temporal correlations. Unlike conventional RNNs, the LSTM introduces a gating mechanism for the explicit regulation of information flow over time, which effectively mitigates the common long-sequence problems of vanishing and exploding gradients (Fig.~\ref{Fig.1}(c)). The introduction of independent memory cells can retain key information over long times, while selectively forgetting past information and controllably writing new information through gated structures \cite{https://doi.org/10.48550/arxiv.1909.09586}. LSTM has the ability to learn the long-term and short-term temporal correlations hidden in the data, thereby can efficiently predict the evolution of open quantum systems, including the complex non-Markovian case.

At a discrete time step $t$, an LSTM unit consists of a hidden state $\mathbf{h}_\mathbf{t}$ and a memory state $\mathbf{c}_\mathbf{t}$. The hidden state represents the output features at the current time step, while the memory state is responsible for storing long-term dependency information during temporal evolution. Their updates are jointly determined by the forget gate, input gate, and output gate, with the evolution equations given by \cite{Krichen2025}
\begin{equation}
	\begin{aligned}
		\begin{bmatrix}
			\mathbf{f}_t \\
			\mathbf{i}_t \\
			\mathbf{o}_t \\
			\widetilde{\mathbf{c}}_t
		\end{bmatrix}
		&=
		\begin{bmatrix}
			\sigma \\
			\sigma \\
			\sigma \\
			\tanh
		\end{bmatrix}
		\!\left(
		W
		\begin{bmatrix}
			\mathbf{x}_t \\
			\mathbf{h}_{t-1}
		\end{bmatrix}
		+
		\mathbf{b}
		\right), \\
		\mathbf{c}_t &= \mathbf{f}_t \odot \mathbf{c}_{t-1}
		+ \mathbf{i}_t \odot \widetilde{\mathbf{c}}_t, \\
		\mathbf{h}_t &= \mathbf{o}_t \odot \tanh\!\left(\mathbf{c}_t\right).
	\end{aligned}
\end{equation}
where $\sigma(\cdot)$ denotes the sigmoid activation function,
$W$ and $\mathbf{b}$ are the concatenated weight matrix and bias vector
corresponding to the forget, input, output, and candidate cell gates,
respectively. $\odot$ denotes element-wise multiplication.

In this work, the LSTM is employed to learn the discrete-time evolution mapping of an open quantum system. At time step $t$, the network input is defined as
\begin{equation}
\mathbf{x}_t = [\langle \sigma_t \rangle, s_t, s_{t+1}, c_t, c_{t+1}, \mathrm{params}(\Gamma,\gamma,T), t],
\end{equation}
where $\langle \sigma_t \rangle = (\langle \sigma_t ^x\rangle, \langle \sigma_t ^y\rangle,\langle \sigma_t ^z\rangle)$ denotes the expectation values of Pauli operators. $(s_t, s_{t+1}, c_t, c_{t+1})$ represent the control amplitudes at the current and next time steps; $\mathrm{param}(\Gamma, \gamma, T)$ denotes the environmental parameters, which are randomly sampled within a certain range in the dataset, including the system--bath coupling strength $\Gamma$, the bath characteristic frequency $\gamma$, and the temperature $T$. $t$ is the time index.

The network performs prediction in an autoregressive manner, with the state update as
\begin{equation}
\mathbf{h}_{t+1}, \mathbf{c}_{t+1} = \mathrm{LSTM}\left(\mathbf{x}_t, \mathbf{h}_t, \mathbf{c}_t\right), \quad 
\langle \sigma_{t+1}^{pred} \rangle = \mathrm{Linear}\left(\mathbf{h}_{t+1}\right).
\end{equation}

Here through a linear layer, the hidden state is mapped to obtain the prediction $\langle \sigma_{t+1}^{pred} \rangle$ at the next time step. This prediction result is then fed back as input for subsequent time steps, enabling step-by-step generation of the system dynamics. During training, the loss function is chosen as the mean squared error (MSE),
\begin{equation}
\mathrm{Loss} = \frac{1}{N} \sum_{t=0}^{N-1} \left\| \langle \sigma_{t+1}^{pred} \rangle - \langle \sigma_{t+1}^{true} \rangle \right\|_F^2,
\end{equation}
where $\|\cdot\|_F$ denotes the Frobenius norm. By minimizing this loss function, the network learns the time-dependent relationships among the driving and control pulses, environmental parameters, and the states of the system. Previous work has shown that training on Gaussian random driving fields is sufficient for the LSTM to learn all the essential features required for predicting the dynamics for arbitrary driving fields \cite{Mohseni2022}, here we use the random driving and control fields to train the NN. And learning is based purely on the expectation values of the observations \cite{An2025}, which are calculated by solving the master equation in Eqs.~\eqref{eq:3}--\eqref{eq:5} with the fourth-order Runge-Kutta method (RK4). That is, the network does not learn the full density matrix $\rho(t)$ directly, but rather the ensemble-averaged dynamics of the Pauli operators. The density matrix $\rho(t)$ can be reconstructed from these expectation values to calculate the fidelity with respect to the target state.

The initial state is first encoded into the LSTM hidden state via a multilayer perceptron (MLP), after which the network autoregressively generates the system evolution trajectory over the entire time interval under given control pulses and environmental parameters (Fig.~\ref{Fig.1}(b)). After training, the network can be used to predict dynamics for arbitrary initial states, environmental parameters, and driving sequences. These generated dynamics can
then be supplied to Adam algorithm that optimizes the pulse shapes to achieve high fidelity, as illustrated in Fig.~\ref{Fig.1}(c).

To enable the network to handle arbitrary time-dependent driving and control pulses, the training samples are generated using Random Fourier Synthesis (RFS) with multi-frequency random sampling. The time-dependent signal is constructed as a superposition of sinusoidal components
\begin{equation}
	s(t)\ \text{or}\ c(t) = \sum_{k=1}^{K} A_k \sin(2\pi f_k t + \phi_k),
\end{equation}
where the amplitudes $A_k$, frequencies $f_k$, and phases $\phi_k$ are randomly sampled from predefined ranges, with the frequencies spanning the dynamical bandwidth relevant to the system. This method ensures broad spectral coverage of the training data, allowing the LSTM to learn the system’s dynamic response under continuous driving and control, enabling robust extrapolation to previously unseen pulse shapes beyond the training distribution.

We first demonstrate the power of the LSTM in predicting the open quantum system dynamics. The network consists of four hidden layers, each with 128 units and tanh activation functions. It is trained on a dataset of 50,000 samples using the Adam optimizer with a learning rate of 0.001 to minimize the mean squared error (MSE) over 200 epochs, with a batch size of 128, ensuring rapid convergence across a broad parameter regime ($\Gamma\in[0,0.05]$, $\gamma\in[1,30]$, $T\in[5,15]$). In Fig.~\ref{Fig.2} we check the prediction accuracy as a function of time. In the first row, we choose three kinds of samples for different $s(t)$ and $c(t)$. They are (i) $s(t)$, $c(t)$ (RFS); (ii) $s(t)=t/T_{tot}$ (linear), $c(t)=I\sin(2\pi t)$ ($I\approx54.4\ \text{to satisfy the third zero of } J_0(I\tau/\pi))$ (sine); (iii) $s(t)=\frac{1}{2}\sin(\pi t/T_{tot}-\frac{\pi}{2})+\frac{1}{2}$ (sine), $c(t)$ ($\tau=2,I\approx37.7 \ \text{to satisfy}\ \ I\tau = 2k\pi\ \text{with}\ \ k=6$) (quench), respectively. The second to the fourth columns are the time evolution of expectation value of $\sigma_{x,(y,z)}$, respectively. The last row corresponds to the MSE, which measures the overall prediction error, providing a measure of the model’s overall prediction accuracy for the quantum state. Throughout the entire time window, the MSE consistently remains below 0.0025 for all three pulse types. The time windows highlighted in light gray and light blue represent the intervals that have and have not been trained on, respectively. As shown in Fig.~\ref{Fig.2}, the evolution curves of the true value and the prediction closely match. This demonstrates that the LSTM model accurately predicts the dynamics across both the training and extrapolation intervals.

\begin{figure}[htbp]
	\centering
	\includegraphics[width=\linewidth]{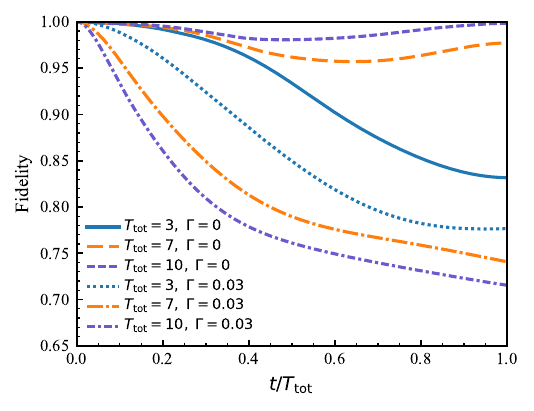}
	\caption{The fidelity $F$ versus the rescaled time $t/T_{tot}$ for different total evolution time $T_{tot}$ without $(\Gamma=0)$ and with $(\Gamma=0.03)$ environment.}
	\label{Fig.3}
\end{figure}

\subsection{Traditional algorithm for the control optimization}

In the presence of environment, the ideal pulse conditions derived in closed cases will lose their effectiveness \cite{Wangzhaoming2020}. In this paper, we use Adam algorithms to design the optimal control.  
Adam is an efficient and scalable gradient-based optimization algorithm that dynamically adjusts the learning rate for each parameter, making it well suited for the present application. To efficiently explore the continuous control space and capture both low- and high-frequency components of the pulses, we parameterize the control functions using Fourier basis expansions \cite{Lin2020}
\begin{equation}
s(t) = \frac{t}{T_{tot}} + \sum_{k=N_s}^{M_s} I_k^s \sin [(k+1) \frac{\omega_s t}{T_{tot}}],
\end{equation}
\begin{equation}
c(t) = \sum_{k=N_c}^{M_c} I_k^c \sin[(k+1) \frac{\omega_c t}{T_{tot}}], 
\end{equation}
where $I_k^{dr}$ are the Fourier coefficients serving as optimization variables and the superscript $dr$ denotes the driving. $dr=s, c$ are the driving through $s(t)$ and $c(t)$, respectively. $\omega_{dr}$ are the fundamental frequencies. $N_{dr}$ $(N_{dr} \geq 1)$ and $M_{dr}$ denote the cutoff values for the low- and high-frequency components, respectively. This parameterization represents the pulse with a finite set of coefficients, where low-frequency terms shape the overall pulse and high-frequency terms adjust fine features. Optimizing these coefficients with Adam efficiently explores the control space to generate smooth, high-fidelity pulse profiles.

The target is to generate optimal pulse profiles by optimizing the coefficients $I_k^{dr}$, thereby enhancing the adiabatic fidelity. This can be obtained by defining the loss function
\begin{equation}
\mathrm{Loss}(I_k^{dr}) = 1 - F(I_k^{dr}) + \lambda dr_{\mathrm{max}},
\end{equation}
where $dr_{max}$ ($dr=s, c$) represents the maximum amplitude of $s(t)$ and $c(t)$. In the loss function, there exists a competition between the fidelity and the maximum pulse amplitude. We introduce a relaxation parameter $\lambda$ to balance their relative weights. Without $\lambda$ term, Adam algorithm tends to achieve maximum fidelity at the expense of continuously increasing control strength, which may be experimentally infeasible \cite{Xie2022}. The detailed description of the Adam algorithm is given in Table I. 

\begin{algorithm}[H]
	\caption{Adam}
	\label{alg:adam}
	\begin{algorithmic}
		\Require Learning rate $\alpha$; exponential decay rates $\beta_1 \in (0,1)$, $\beta_2 \in (0,1)$; numerical stability constant $\varepsilon$; initial control parameters (Fourier coefficients) $I^{(0)} = I_k^{dr}$; maximum number of iterations $K_{\max}$.
		\Ensure Optimized control parameters $I^*$
		\State Set first moment vector $\mathbf{m}_0 = \mathbf{0}$, second moment vector $\mathbf{v}_0 = \mathbf{0}$, iteration counter $K = 0$.
		\While{$K < K_{\max}$}
		\State Evaluate loss: $\mathrm{Loss}(I^{(t)}) = 1 - F(I^{(t)}) + \lambda (s_{\max} \ \text{or}\ c_{\max})$
		\State Compute gradient: $\mathbf{g}_t = \nabla_I \mathrm{Loss}(I^{(t)})$
		\State Update biased first and second moment estimates:
		\State \quad $\mathbf{m}_t = \beta_1 \mathbf{m}_{t-1} + (1 - \beta_1) \mathbf{g}_t$
		\State \quad $\mathbf{v}_t = \beta_2 \mathbf{v}_{t-1} + (1 - \beta_2) \mathbf{g}_t^2$
		\State Compute bias-corrected estimates:
		\State \quad $\hat{\mathbf{m}}_t = \mathbf{m}_t / (1 - \beta_1^t)$
		\State \quad $\hat{\mathbf{v}}_t = \mathbf{v}_t / (1 - \beta_2^t)$
		\State Update parameters: $I^{(t+1)} = I^{(t)} - \alpha \frac{\hat{\mathbf{m}}_t}{\sqrt{\hat{\mathbf{v}}_t} + \varepsilon}$
		\State $K \gets K + 1$
		\EndWhile
		\State \Return $I^* = I^{(t)}$
	\end{algorithmic}
\end{algorithm}

\begin{figure}[htbp]
	\centering    
	\includegraphics[width=\linewidth]{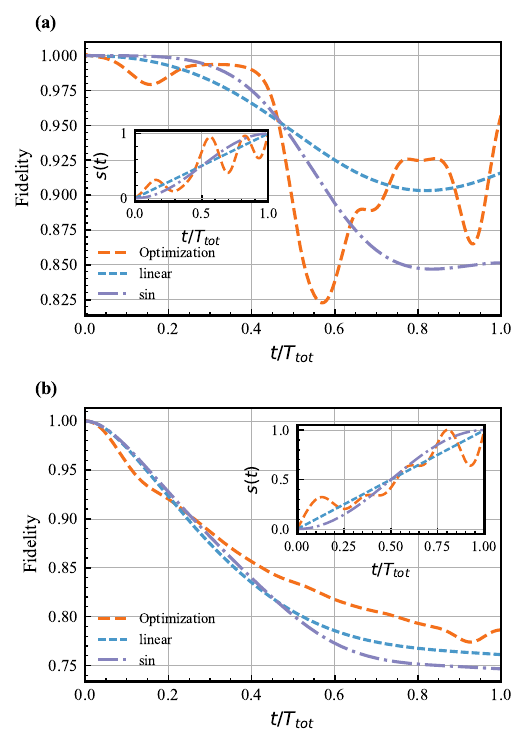}
	\caption{The fidelity versus the rescaled time $t/T_{tot}$ for linear ($s(t)=t/T_{tot}$), sine ($s(t)=\frac{1}{2}\sin(\pi t/T_{tot}-\frac{\pi}{2})+\frac{1}{2}$)}, and optimal $s(t)$ without (Fig.~\ref{Fig.4}(a)) and with (Fig.~\ref{Fig.4}(b), $\Gamma=0.03, \gamma=2, T=10$) environment. $T_{tot}=5$.
	\label{Fig.4}
\end{figure}

\begin{figure}[htbp]
	\centering
	\includegraphics[width=\linewidth]{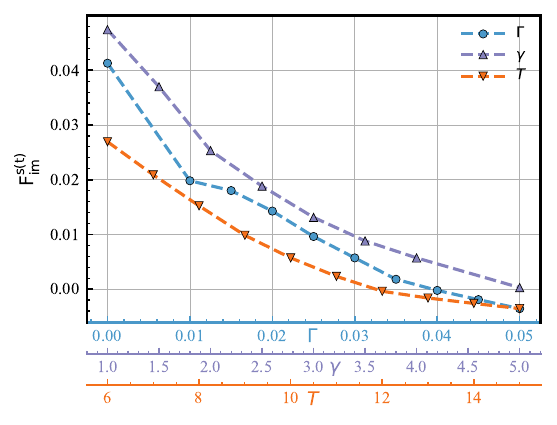}
	\caption{The fidelity improvement $F_{im}^{s(t)}$ versus environmental parameters $\Gamma, \gamma$ and $T$. $F_{im}^{s(t)}=F_{opt}^{s(t)}-F_{lin}^{s(t)}$, with $F_{opt}^{s(t)}$ and $F_{lin}^{s(t)}$ represent the final fidelities corresponding to the optimized and the linear $s(t)$, respectively. $\Gamma=0.03, \gamma=4, T=5$, and the other two are fixed when one of them varies.}
	\label{Fig.5}
\end{figure}

\begin{figure*}[htbp]
	\centering
	\includegraphics[width=\textwidth]{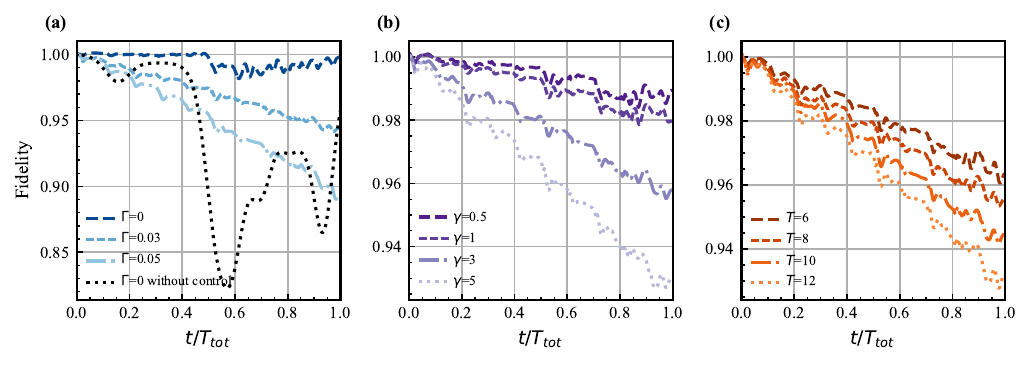}
	\caption{The fidelity versus the rescaled time $t/T_{tot}$ under ideal pulse control $c(t)=$ for different (a) $\Gamma$; (b) $\gamma$ and (c) $T$. $\Gamma=0.03, \gamma=4$, $T=10$, $T_{tot}=5$ and the other two parameters are fixed when one parameter varies. }
	\label{Fig.6}
\end{figure*}

\section{Results and Discussions}
Now with the LSTM predicted dynamics, we first identify the approximate adiabatic regime. In Fig.~\ref{Fig.3} we plot the total evolution time $T_{tot}=3,7,10$ without ($\Gamma=0$) and with $(\Gamma=0.03)$ environment, respectively. The other environmental parameters are chosen to be $\gamma=2, T=10$. At first without environment, the adiabatic fidelity $F$ increases with increasing $T_{tot}$, and with $T_{tot}\approx 10$, the system enters the adiabatic regime. In the presence of environment, the fidelity will be destroyed by the environment. In this case, $F$ will not always increase with increasing $T_{tot}$. It can be seen that for a short time  $T_{tot} = 3$, the environment does not affect the fidelity very much. With a longer $T_{tot} =7$, it begins to reduce the fidelity. For $T_{tot} =10$, which is in an adiabatic regime without environment, the environment reduces the fidelity even more. The reason is that the detrimental effects of the environment will accumulate with increasing $T_{tot}$, which has been observed in Ref.~\cite{Wang2018adiabatic}. We emphasize that the difference is that in Ref.~\cite{Wang2018adiabatic} the dynamics are obtained through numerical calculation, whereas they are predicted by the LSTM in this study.

Next we turn to the optimization problems with our LSTM predicted dynamics. We take two-step optimization strategy for improving the adiabaticity. The first step is the optimization of the driving trajectory $s(t)$. The choice of driving trajectory $s(t)$ determines the performance of the adiabatic algorithm and the complexity scaling. For instance, in Grover search, a linear $s(t)$ leads to $O(N)$ in the search space, while it becomes $O(\sqrt{N})$ for a nonlinear choice of $s(t)$ \cite{Roland2002}.  Then the design of the $s(t)$ is an important task in the adiabatic quantum computing. Optimization of $s(t)$ using reinforcement learning \cite{Lin2020} or the Gradient Ascent Pulse Engineering algorithm \cite{https://doi.org/10.48550/arxiv.2506.12138} has been studied recently.

Fig.~\ref{Fig.4} plot the fidelity versus the rescaled time $t/T_{tot}$ for linear, sine and optimal $s(t)$ without (Fig.~\ref{Fig.4}(a)) and with (Fig.~\ref{Fig.4}(b), $\Gamma=0.03, \gamma=2, T=10$) environment. $T_{tot}=5$. The sine trajectory outperforms the linear one in the first half of the total duration $T_{tot}$, while the linear trajectory becomes superior in the second half. This relative performance holds true both with and without environmental factors. Notably, the optimal trajectory demonstrates its advantage at the final time $t=T_{tot}$ in all cases. The inset of Fig.~\ref{Fig.4}(a) and (b) plots the corresponding $s(t)$. During the optimization, we have limited the value of $s(t)$ in the regime $[0,1]$ with $s(0)=0$ and $s(t=T_{tot})=1$.

As shown in Fig.~\ref{Fig.4}(b), optimizing $s(t)$ alone yields only a modest fidelity improvement with certain environmental parameters. To demonstrate the fidelity improvement with environmental parameters, we define the fidelity improvement as $F_{im}^{s(t)}=F_{opt}^{s(t)}-F_{lin}^{s(t)}$, with $F_{opt}^{s(t)}$ and $F_{lin}^{s(t)}$ represent the final fidelities corresponding to the optimized and the linear $s(t)$, respectively. We take $\Gamma=0.03, \gamma=4, T=5$, and fix the other two when one of them varies. Fig.~\ref{Fig.5} demonstrate that $F_{im}^{s(t)}$ gradually decreases with increasing all these three parameters. This indicates that in a stronger environment, it becomes increasingly difficult to enhance the adiabatic fidelity through the optimization of $s(t)$.

To further improve the adiabatic fidelity, we employ a global zero-area pulse control scheme. This method accelerates the adiabatic evolution even in a non-adiabatic regime with a short $T_{tot}=5$, thereby effectively mitigating the impact of noise. In this case, the system's exposure time to the environment is reduced and thus the noise accumulation is limited. We first use the ideal pulse control which obtained in closed system. Fig.~\ref{Fig.6} plots the fidelity evolution under idea pulse control with different $\Gamma$ (Fig.~\ref{Fig.6}(a)), $\gamma$ (Fig.~\ref{Fig.6}(b)) and $T$ (Fig.~\ref{Fig.6}(c)), respectively. $\Gamma=0.03, \gamma=4$, $T=10$, $T_{tot}=5$ and the other two parameters are fixed when one parameter varies. From Fig.~\ref{Fig.6}(a), without control the fidelity will decrease quickly even without environment ($\Gamma=0$). Under the ideal pulse $c(t)=I\sin(2\pi t)$, the fidelity will be significantly improved with and without environment. Here $I\approx54.4$, satisfying the third zero of $J_0(I\tau/\pi)$. However, the ideal pulse control conditions
are only applicable in closed systems and lose their
effectiveness in open systems. Clearly, the fidelity decreases with increasing parameters $\Gamma$, $\gamma$ and $T$, indicating that as the environmental influence intensifies, the control performance deteriorates.

\begin{figure}[htbp]
	\centering
	\includegraphics[width=\linewidth]{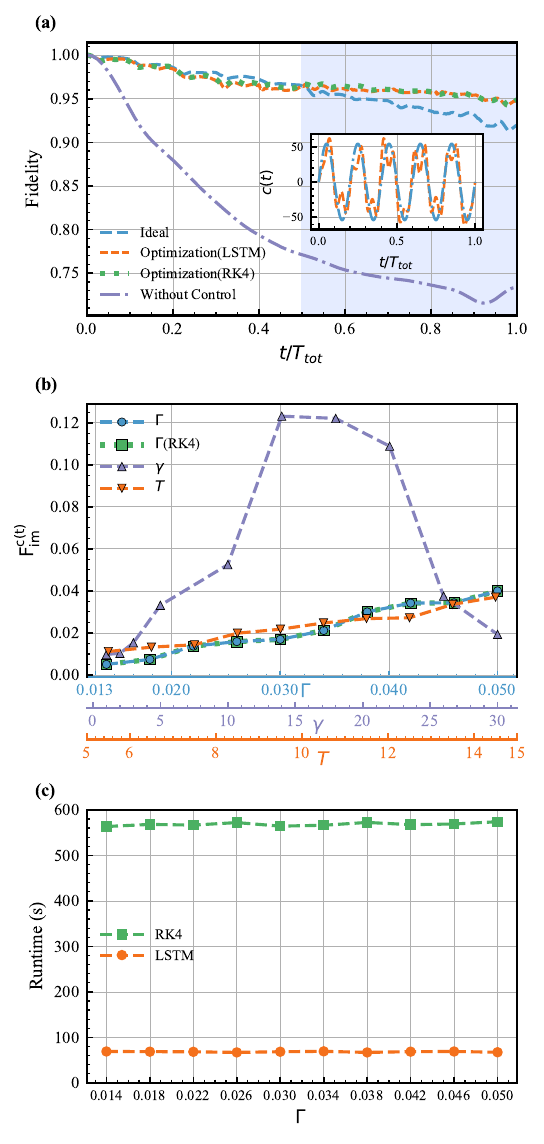}
	\caption{(a) The fidelity versus the rescaled time $t/T_{tot}$ without, with ideal, and optimized control  where  $\Gamma=0.04$, $\gamma=4$, $T=10$, The ideal control $c(t)=I\sin(2\pi t)$($I\approx54.4\ \text{to satisfy the third zero of } J_0(I\tau/\pi))$. The inset shows the corresponding pulse. The time window $[2.5, 5]$ highlighted in light purple corresponds to the region where the NN are not trained; (b) The fidelity improvement $F_{im}^{c(t)}$ versus environmental parameters $\Gamma, \gamma$ and $T$. $F_{im}^{c(t)}=F_{opt}^{c(t)}-F_{ideal}^{c(t)}$, with $F_{opt}^{c(t)}$ and $F_{ideal}^{c(t)}$ represent the final fidelities corresponding to the optimized and the ideal $c(t)$, respectively. $\Gamma=0.03, \gamma=4, T=10$, and the other two are fixed when one of them varies; (c) Runtime versus environmental parameter $\Gamma$, comparing the time needed for the pulse design via the NN against direct numerical calculation.}
	\label{Fig.7}
\end{figure}

We then take the second step: optimization of the zero-area pulse control function $c(t)$. A central objective is to partially mitigate environmental disruption to adiabaticity by adapting the ideal pulse shape originally designed for closed systems, thereby identifying a noise-insensitive evolution path.
Fig.~\ref{Fig.7}(a) plots the fidelity as a function of the rescaled time $T_{tot}$ without pulse control, with ideal and optimal pulses control, respectively. The profiles of the corresponding ideal and optimal pulses are demonstrated in the inset of  Fig.~\ref{Fig.7}(a). The environmental parameters are taken as $\Gamma=0.04$, $\gamma=4$, $T=10$. The performance of the optimal $c(t)$ are better than the ideal ones, which certainly also outweigh the free evolution case. 
Similar with Fig.~\ref{Fig.5}, we also define a fidelity improvement as $F_{im}^{c(t)}=F_{opt}^{c(t)}-F_{ideal}^{c(t)}$, with $F_{opt}^{c(t)}$ and $F_{ideal}^{c(t)}$ represent the final fidelities corresponding to the optimized and the ideal $c(t)$, respectively. We take $\Gamma=0.03, \gamma=4, T=10$, and fix the other two when one of them varies. Fig.~\ref{Fig.7}(b) demonstrate similar behavior as in Fig.~\ref{Fig.5}, $F_{im}^{c(t)}$ gradually increases with increasing parameters $\Gamma$ and $T$. This trend supports the physical intuition that a stronger environment (larger \(\Gamma\), higher $T$) disrupts adiabaticity more severely, thus offering greater scope for enhancing fidelity through optimized pulses. However,  $F_{im}^{c(t)}$ does not change monotonically with increasing \(\gamma\), a peak exists at a critical value (\(\gamma\approx15\)). This result is in accordance with Refs. \cite{xie2022stochastic,tai2014optimal,wang2021quantum}: A higher \(\gamma\) yields a more Markovian and disruptive environment, which tends to raise $F_{im}^{c(t)}$. However, this tendency is weakened by a reduction in the control’s effectiveness. At last, to verify the accuracy of the prediction, in Fig.~\ref{Fig.7}(a),(b) we also plot the direct numerical calculations of the non-Markovian master equation (RK4) for optimal pulse case. The light blue region of Fig.~\ref{Fig.7}(a) corresponds to the time interval that the NN is not trained. The fidelity evolution for the optimized case in Fig.~\ref{Fig.7}(a) closely matches between the LSTM and RK4 methods, whether the network is trained or not.

In Fig.~\ref{Fig.7}(b), we compare the $\Gamma$ depedence of the fidelity for LSTM predicted dynamics and RK4 calculated dynamics. As expected the two results are almost identical, showing the high accuracy of the LSTM in predicting the dynamics. To show the advantage of our strategy,  in Fig.~\ref{Fig.7}(c) shows the runtime for parallel optimization of 64 control sequences over 200 iterations, comparing the LSTM-based approach with direct numerical calculation using RK4. For a fair comparison, both methods employ the same time step and all simulations were performed on a single NVIDIA GTX 1650 GPU. As illustrated, for different parameter $\Gamma$,  the LSTM require approximately 60 s, whereas the RK4 evolution takes about 580 s, corresponding to a speedup of roughly one order of magnitude. It should be noted that the reported LSTM runtime does not include the offline training time of the network. However, once trained, the LSTM can predict the dynamics for a wide range of different driving pulses and parameter settings, providing a significant overall computational advantage. This speedup originates from differences in the computational structure and the use of local observables. The RK4 propagation requires four matrix evaluations per time step with repeated Hamiltonian applications, leading to significant overhead. In contrast, the LSTM relies on fixed tensor operations with highly optimized GPU execution. Moreover, while both approaches involve gradient-based optimization, the LSTM benefits from efficient automatic differentiation and low-level kernel optimization in PyTorch, whereas RK4-based control typically requires repeated forward simulations or adjoint constructions. In addition, the LSTM predicts only selected observables rather than the full quantum state, further reducing computational cost. Despite this limitation, for typical quantum dynamical tasks such as trajectory prediction and control pulse optimization, the LSTM achieves a speedup of one to two orders of magnitude while maintaining accurate predictions of the target observables.

\begin{figure}[htbp]
	\centering
	\includegraphics[width=\linewidth]{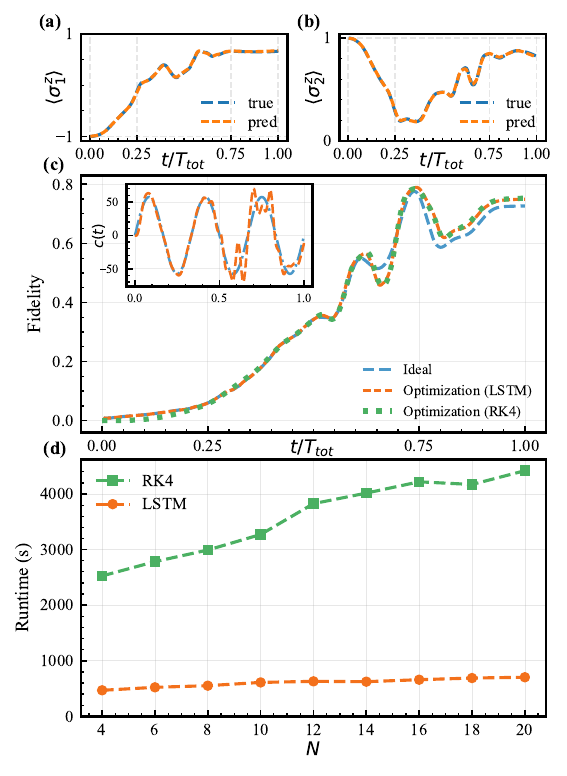}
	\caption{(a)–(b) The local observables $\langle \sigma_1^z \rangle$ and $\langle \sigma_2^z \rangle$ versus the rescaled time $t/T_{tot}$; (c) The transmission fidelity versus the rescaled time $t/T_{tot}$ under ideal, and optimized control. The inset shows the corresponding pulse;	(d) Comparison of pulse design runtime as a function of chain length $N$ using LSTM versus direct numerical calculation. $T_{tot}=\pi/4$, $\Gamma=0.3$, $\gamma=10$, $T=15$, the ideal control $c(t)=I\sin(2\pi t)$($I\approx57.7\ \text{to satisfy the first zero of } J_0(I\tau/\pi))$. $N=4$ in (a)-(c).}
	\label{Fig.8}
\end{figure}

To further test scalability beyond one qubit system, next we apply our algorithm to a high-dimensional system. As an example, we study quantum state transfer through an $XY$ spin chain which is immersed in noisy environments.
The system Hamiltonian reads
\begin{eqnarray}
	H_s = \sum_{i=1}^{N-1} J_{i,i+1}
	\left(\sigma_i^x \sigma_{i+1}^x + \sigma_i^y \sigma_{i+1}^y \right),
\end{eqnarray}
where $J_{i,i+1}$ denotes the nearest-neighbor coupling strength between sites $i$ and $i+1$. $N$ is the length of the chain. We choose $J_{i,i+1}=\sqrt{i(N-i)}$, which enables perfect state transfer (PST) in the absence of the environment ~\cite{christandl2004perfect}. We take the model that two end spins of the chain are immersed in its environment \cite{ren2019quantum}, and the Lindblad operator is taken as $L=\sigma_1^z+\sigma_N^z$. In this case, the dynamics can be described within the single-excitation subspace. The simple state transfer task $|\mathbf{1}\rangle \rightarrow |\mathbf{N}\rangle$ is considered here, where $|\mathbf{i}\rangle$ denotes the state with the $i$-th spin excited and all others in the ground state. With leakage elimination control, the Hamiltonian becomes: $H = H_{\mathrm{s}} + H_{\mathrm{LEO}},$ where $H_{\mathrm{LEO}} = c(t)|\psi_0(t)\rangle \langle \psi_0(t)|$ with $|\psi_0(t)\rangle = e^{-iH_{\mathrm{PST}} t} |\mathbf{1}\rangle$, and $c(t)$ is the control pulse to be optimized~\cite{Wang2020pulsecontrol}. Here, the ideal control is given by
$c(t)=I\sin(2\pi t)$, where $I\approx57.7$ is chosen to satisfy the first zero of $J_0(I\tau/\pi)$.
The local observables are chosen as $\langle\sigma_z^i \rangle$ ($i=0,1,\dots,N$).

Figure~\ref{Fig.8} illustrates the performance of the proposed method in the $4$-site spin chain. Figures~\ref{Fig.8}(a)–(b) show the time evolution of local observables $\langle \sigma_z^1 \rangle$ and $\langle \sigma_z^2 \rangle$ under the optimized control protocol, with parameters $T_{tot}=\pi/4$, $\Gamma=0.3$, $\gamma=10$, and $T=15$. The LSTM-predicted dynamics agree well with the optimized results, indicating reliable generalization to high-dimensional systems. Figure~\ref{Fig.8}(c) presents the time evolution of the fidelity  with the same parameters as in Figures~\ref{Fig.8}(a)–(b), comparing the ideal and optimized control protocols, together with a comparison between LSTM and RK4 calculations. The optimized control enhances the fidelity, demonstrating the effectiveness of the algorithm in high-dimensional systems. The corresponding ideal and optimized control pulses are shown in the inset of Fig.~\ref{Fig.8}(c). Figure~\ref{Fig.8}(d) plots the runtime required for optimizing a single set of control parameters over 200 Adam iterations as a function of the chain length $N$. To clearly characterize the scaling behavior with system size, all calculations are performed on Intel X7 358H CPU under a fixed single-thread setting, in order to eliminate acceleration effects from GPU parallelization. The results show that, as $N$ increases from 4 to 20, the runtime of the RK4 method increases from approximately 2500 s to 4300 s, whereas the LSTM-based method increases only from approximately 450 s to 700 s. The computational cost of RK4 grows rapidly with system size in the single-excitation subspace due to the propagation of \(N \times N\) density matrices and repeated matrix operations at each time step. In contrast, the LSTM relies only on local observable representations, leading to an approximately linear scaling with $N$. As a result, the LSTM achieves significantly lower computational cost while maintaining high accuracy, exhibiting favorable scaling behavior with system size and highlighting the scalability of the algorithm for high-level system quantum optimal control.

\section{CONCLUSIONS}
In this paper, we have introduced LSTM for the prediction of open quantum system dynamics into the field of quantum control. Through training the NN on numerical data solved from the non-Markovian master equation, the NN learns the dynamics of the interested observables. Leveraging this ability, integrating traditional optimal algorithm like Adam, an optimal control design can be obtained in a shorter time than direct numerical calculation. We first apply this strategy to adiabatic speedup in a two level system under a non-Markovian environment. A two step optimization is employed to improve the adiabatic fidelity, including optimization of both the driving trajectory and the external control. We then extend the strategy to high-dimensional control problems, specifically to quantum state transfer in a spin chain. The results demonstrate high consistency between the predicted outcomes and those from direct numerical calculation, and significant fidelity improvement is achieved.

Our findings open avenues for applying LSTM to quantum control, demonstrating that combining LSTM-predicted dynamics with traditional optimal control algorithms can significantly reduce design time.  The approach is particularly powerful for processing real experimental data, as it can operate using only partial observables of the quantum system. Future work could extend this method to more complex tasks, such as optimal pulse design for quantum error correction or practical quantum computation on platforms like superconducting quantum processors \cite{Ge2026}.

\section*{acknowledgements}
We thank Zongyuan Ge for helpful discussions on the LSTM. This paper is based upon work supported by the  Natural Science Foundation of Shandong Province (Grant No. ZR2024MA046).

{\section*{DATA AVAILABILITY} The data are available from the corresponding author upon reasonable request.
}

\bibliography{ref.bib}

\end{document}